# Altermagnetism-Induced Spin-Resolved Electronic Structures in Janus $FeX_{0.5}Y_{0.5}$ Monolayers (X, Y = S, Se, Te)

Mengyang Zhang, Jie Li,* Yifei Chen, Shifang Li, Zhentao Fu, and Jianxin Zhong*

Institute for Quantum Science and Technology, Shanghai University, Shanghai 200444, China

Realizing the spin-resolved electronic properties in superconducting materials stands as a critical frontier, offering both novel fundamental physics and potential for dissipationless spin-based devices. Here, we predict a series of Janus $FeX_{0.5}Y_{0.5}$ monolayers derived from iron-based superconductors (e.g., FeSe, FeTe, and FeS) by using Kondo-type model and first-principles calculations. These Janus structures exhibit significant spin-splitting electronic states, large topological band gaps (51.4 meV) and high Néel temperatures (415 K). We further reveal that valley polarization can be effectively tuned via applied in-plane strain and the resulting valley-polarized anomalous Hall conductivity can be manipulated by shifting the Fermi level. Our work suggests a new strategy based on altermagnetism for engineering spin-splitting states in superconducting systems and inspires further exploration of superconducting spintronics.

*Corresponding authors: lij@shu.edu.cn and jxzhong@shu.edu.cn

## I. INTRODUCTION

Superconductivity stands as one of the most central topics of modern condensed matter physics and quantum information science, drawing significant experimental and theoretical attention. Traditionally, most investigations have been focused on their transition temperatures and pairing mechanisms. In recent years, topological superconducting states haven been received considerable attention in a variety ststems including Kitaev quantum wires [1], topological superconducting heterostructures [2], semiconductor–superconductor heterostructures [3-5], ferromagnetic atomic chains on superconductors [6], and iron-based superconductors [7,8]. Realization of spin-resolved electronic properties in superconducors has also emerged as a critical frontier, offering both novel fundamental physics and potential for dissipationless spin-based devices [9-11]. Over the past decades, several routes have been developed to generate, manipulate, and detect spin polarization in superconducting systems, from the magnetic proximity effect at superconductor/ferromagnet interfaces [9,12], combination of strong spin–orbit coupling and local inversion symmetry breaking [13], heavy-fermion superconductors [14] to the conversion of spin-singlet Cooper pairs into spin-triplet pairs in ferromagnetic Josephson junctions [15]. However, conventional ferromagnetic order is generally incompatible with superconductivity, particularly in the spin-singlet channel, as the exchange field polarizes the electronic bands and creates an energy offset between electrons of opposite spins, thereby inhibiting conventional Cooper pair formation. Furthermore, the stability of the coexisting superconducting and spin-polarized phases is critically dependent on both the intrinsic quality of the materials and the atomic-scale sharpness of the interfaces. These motivate the exploration of spin-resolved electronic states for superconducting spintronics.

Recently, altermagnets have been proposed as a third class of magnetic materials, characterized by momentum-dependent spin splitting and zero net magnetization. Unlike conventional ferromagnets, this spin splitting does not rely on strong spin-orbit coupling, but rather on the breaking of combined spatial and time-reversal symmetries [16-19]. Such properties naturally motivate the exploration of alternative magnetic orders in superconducting materials that can induce spin-polarized band structures without introducing a net magnetic moment. The similar opposing spin sublattices makes it possible to design altermagnets starting from conventional antiferromagnets [20,21]. In this context, monolayer FeSe and related iron-based superconductors, already possesses an antiferromagnetic ground state. It is well known that monolayer FeSe is a high-temperature superconductor [22-28]. Experiments have also confirmed that the superconducting temperatures can reach 13.5 K for FeTe [29,30] and 4 ~ 5 K for FeS [31-33]. Considering the abundant superconducting peoperies and the significant impact in the research field of topological superconductors, developing a straightforward and robust route to induce altermagnetism in monolayer FeSe related materials would provide a promising platform for accessing spin-splitting electronic states in a superconducting and topological environment.

In this work, we employ iron-based monolayer superconductors (e.g., FeSe, FeTe, and FeS) as a platform and propose a strategy to realize altermagnetism, thereby inducing spin-splitting electronic states in superconducting materials, by introducing anisotropic local environments for the two magnetic sublattices. Through Density Functional Theory (DFT) calculations, monolayers Janus $FeX_{0.5}Y_{0.5}$ (X≠Y; X, Y = S, Se, Te) are shown to be particularly promising platforms, exhibiting desired spin splitting across the entire Brillouin zone, with tunable valley polarization, large topological band gap and high Néel temperature. These findings can enrich realizing and controlling spin-splitting electronic states in superconducting materials, and inspire further explorations on superconducting spintronics.

## II. METHODOLOGY

All the density functional theory (DFT) calculations of this work are carried out with the Vienna ab initio simulation package (VASP) [34]. For the exchange-correlation functional, the generalized-gradient approximation (GGA) developed by Perdew-Burke-Ernzerhof is adopted [35,36]. The Hubbard U = 1.0 eV is used for the electron correlation in the d-shells of Fe cores [37]. In all DFT calculations, the energy cutoff for plane-wave representation is set to 450 eV, and the self-consistent convergence is enforced with thresholds of $10^{-5}$ eV for total energy and 0.01 eV/Å for atomic forces.

## III. RESULTS AND DISCUSSION

To explore the possibility of designing altermagnetism in iron-based superconductors (FeSe, FeTe, and FeS), we propose a duplex checkerboard (DCB) lattice model as shown in Fig. 1(a) composed of a checkerboard lattice formed by two magnetic superlattices (A and B) with antiparallel magnetic moments and a checkerboard lattice formed by two nonmagnetic superlattice (C and D), respectively. The magnetic A and B superlattices are of fourfold rotation symmetry and the nonmagnetic C and D superlattices are alternately embedded into the central hollow sites of the magnetic checkboard lattice. Our DCB-lattice model is an extension of the Lieb-lattice model for altermagnetism [38], in which only one nonmagnetic superlattice (C or D) is considered. As in our previous work [39] for the FeSe bilayer systems, the crystal environment provided by nonmagnetic sites is necessary for the presence of altermagnetism, the nonmagnetic sites are therefore included in our DCB-lattice to tune its magnetic properties in the framework of FeSe related monolayer systems under the Kondo-type model .

Without the spin–orbit coupling (SOC) included, the Hamiltonian of the DCB lattice is derived following the similar approach in Ref. [38]:

$$H_0(\mathrm{k}) = \begin{bmatrix} H_1(\mathrm{k}) & 0 \\ 0 & H_2(\mathrm{k}) \end{bmatrix}, \tag{1}$$

$$H_1(k) = \begin{bmatrix} -2t_2\left(\cos k_x + \cos k_y\right) + JN + \Delta & 4t_1 \cos\frac{k_x}{2}\cos\frac{k_y}{2} \\ -4t_1 \cos\frac{k_x}{2}\cos\frac{k_y}{2} & -2t_2\left(\cos k_x + \cos k_y\right) - JN - \Delta \end{bmatrix}, \tag{2}$$

$$H_2(k) = \begin{bmatrix} -2t_2\left(\cos k_x + \cos k_y\right) - JN + \Delta & -4t_1 \cos\frac{k_x}{2}\cos\frac{k_y}{2} \\ -4t_1 \cos\frac{k_x}{2}\cos\frac{k_y}{2} & -2t_2\left(\cos k_x + \cos k_y\right) + JN - \Delta \end{bmatrix}, \tag{3}$$

where $t_1$ represents the nearest-neighbor hopping, $t_{2a}$, and $t_{2b}$ are the next-nearest-neighbor hoppings with $t_2 = (t_{2a} + t_{2b})/2$, J is the Kondo coupling to the local moments, and $N = M_A - M_B$ is the staggered magnetization. The schematic of the DCB-lattice shown in Fig. 1(a) illustrates that the next-nearest-neighbor hopping between magnetic atoms originates from superexchange mediated by the intervening nonmagnetic atoms ( C or D atoms). Thus, anisotropic next-nearest-neighbor hopping ($t_{2a}$ and $t_{2b}$) can be induced by different crystallographic environments along x- and y-direction. The anisotropy of the next-nearest-neighbor hopping is defined as $\Delta = t_{2a} - t_{2b}$. The band structures of two types of the DCB-lattices, i.e., the isotropic one ($\Delta = 0$) and anisotropic one ($\Delta \neq 0$), are shown in Fig. 1(b) and 1(d), respectively. One can see that the band structure with $\Delta = 0$ exhibits spin-degeneracy throughout the Brillouin zone, with its antiferromagnetism protected by PT-symmetry, while the case of $\Delta \neq 0$ exhibits the desired spin splitting. Consequently, introduction of anisotropic local environments for the two magnetic sublattices is key to achieving the transition from an antiferromagnet to an altermagnet. We note that various methods can be used to induce the anisotropic local environments for the two magnetic sublattices, including interlayer stacking [40,41], external electric fields [42,43], uniaxial strain [44,45], and interlayer magnetoelectric coupling [46,47].

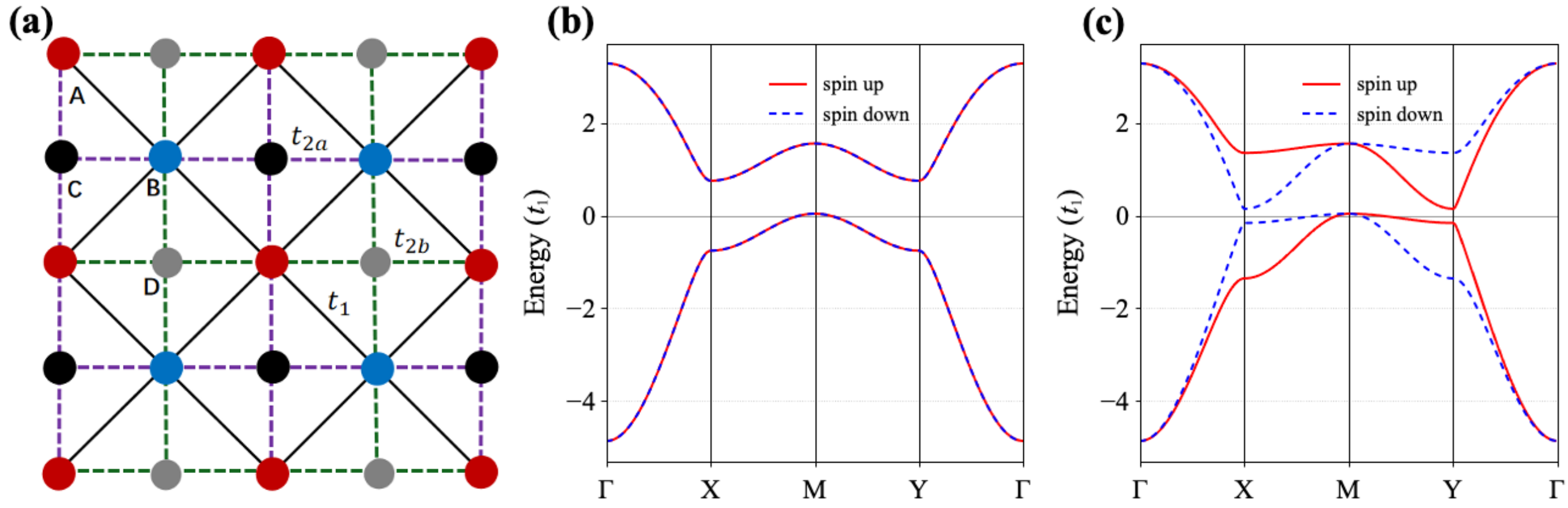

*Fig. 1. (a) Schematic of the DCB-lattice model (red and blue spheres represent atoms in the two antiparallel magnetic sublattices A and B, black and gray spheres represent atoms in the the nonmagnetic superlattices C and D), where* $t_1$ *is the nearest-neighbor hopping between magnetic atoms,* $t_{2a}$ *and* $t_{2b}$ *are the next-nearest-neighbor hoppings between magnetic atoms mediated by the nonmagnetic C and D atoms, respectively. (b) The band structure from the Kondo-type model for* $t_1$ *= 1,* $t_2$ *= 0.2, and* $\Delta$ *= 0. (c) The corresponding band structure for* $\Delta$ *= 0.3. Parameters of* $t_1$*,* $t_2$*, and* $\Delta$ *are in unit of* $t_1$*.*

Guided by the above model analysis, symmetry engineering is adopted to introduce the anisotropic local environments for the two magnetic sublattices in iron-based monlayer superconductors. We consider the Janus structures of $FeX_{0.5}Y_{0.5}$ (X≠Y; X, Y = S, Se, Te) with different chalcogen atoms for the top and bottom atomic layers as shown in Fig. 2(a) and Fig. S1. In this monolayer Janus $FeX_{0.5}Y_{0.5}$, the Bader charge analysis shows that the top layer loses 0.19 e (0.5 e, 0.3 e)/supercell to the bottom layer, which induces a dipole moment of 0.21 eÅ (0.19 eÅ, 0.15 eÅ) for $FeSe_{0.5}Te_{0.5}$ ($FeS_{0.5}Se_{0.5}$, $FeS_{0.5}Te_{0.5}$). As a result, a perpendicular electrostatic potential difference of 0.26 eV (0.27 eV, 0.59 eV) between the top and bottom atomic layers is present in $FeSe_{0.5}Te_{0.5}$, ($FeS_{0.5}Se_{0.5}$, $FeS_{0.5}Te_{0.5}$) as shown in Fig. 2(b) (Fig. S2). DFT calculations show that the optimized $FeSe_{0.5}Te_{0.5}$, $FeS_{0.5}Se_{0.5}$ and $FeS_{0.5}Te_{0.5}$, have in-plane lattice constants of 3.85 Å, 3.72 Å, and 3.79 Å, respectively, which are small different from those of FeSe, FeTe and FeS [48-50].

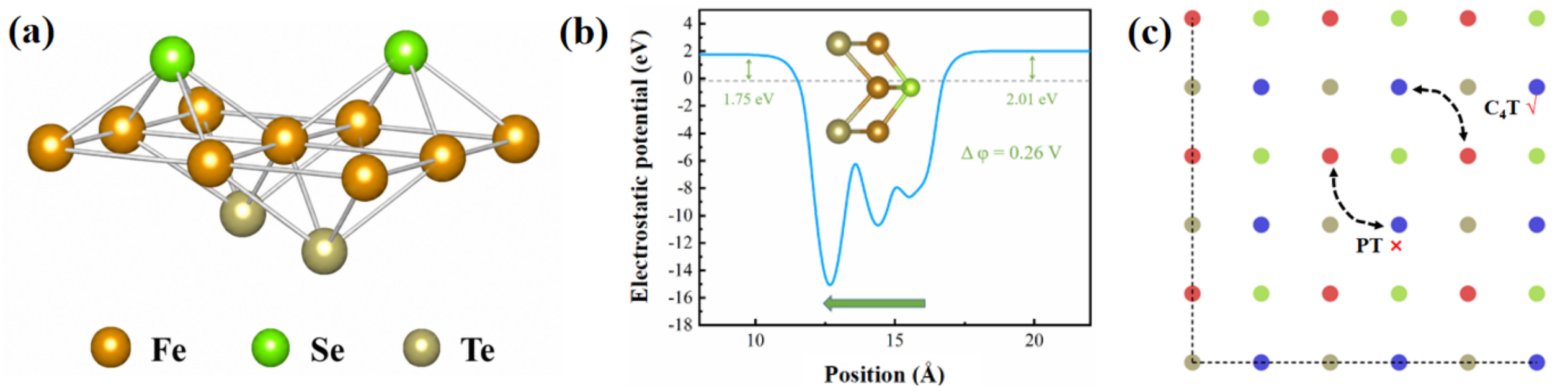

*Fig. 2. (a) Structure of the optimized monolayer Janus $FeSe_{0.5}Te_{0.5}$. (b) The planar-averaged electrostatic potential of the monolayer $FeSe_{0.5}Te_{0.5}$ along z-direction. (c) Schematic of opposite spin sublattices in monolayer $FeSe_{0.5}Te_{0.5}$.*

To confirm the thermal and dynamic stability of $FeX_{0.5}Y_{0.5}$, the corresponding phonon spectra and molecular dynamics simulations are further explored as shown in Fig. S3. The absence of an imaginary frequency branch in phonon bands indicates that these systems are dynamically stable. The ab initio molecular dynamics (AIMD) simulations show that no noticeable structural deformation for a 3×3 supercell after 10000 MD steps (10 ps) at 300 K and the total energies fluctuate around their equilibrium values without noticeable sudden change (see Fig. S3), confirming that these monolayer Janus $FeX_{0.5}Y_{0.5}$ structures are thermally stable at least up to room temperature.

DFT calculations further indicate that $FeSe_{0.5}Te_{0.5}$ ($FeS_{0.5}Te_{0.5}$, $FeS_{0.5}Se_{0.5}$) with the opposing spin orders as shown in Fig. 2(c). From the view of symmetry analysis, the unique Janus structure breaks the corresponding spatial inversion symmetry (P), their spin-space symmetry changes from PT-symmetry to $C_4$T-symmetry after the spin degrees of freedom is included. This symmetry reduction can give rise to the anisotropic local environments for the two magnetic sublattices, which in turn induces altermagnetism, as confirmed by their band structures (see Fig. 3(a), Fig. S6(a) and Fig. S7(a), respectively). There is significant spin splitting across the entire Brillouin zone, particularly in the valence and conduction bands near the Fermi level. The corresponding projected density of states (PDOS) shown in Fig. 3(b) reveals that the spin splitting near the Fermi level predominantly originates from Fe $d_{z^2}$ orbitals. The superexchange interactions between the next-nearest-neighbor Fe atoms along the x-axis and y-axis, corresponding to the next-nearest-neighbor hoppings $t_{2a}$ and $t_{2b}$ as described in the above Kondo-type model, are mediated via the $p_z$ orbitals of Se and Te atoms, respectively. The different superexchange interactions along x-axis and y-axis lead to the desired spin splitting as discussed in the above Kondo-type model analysis.

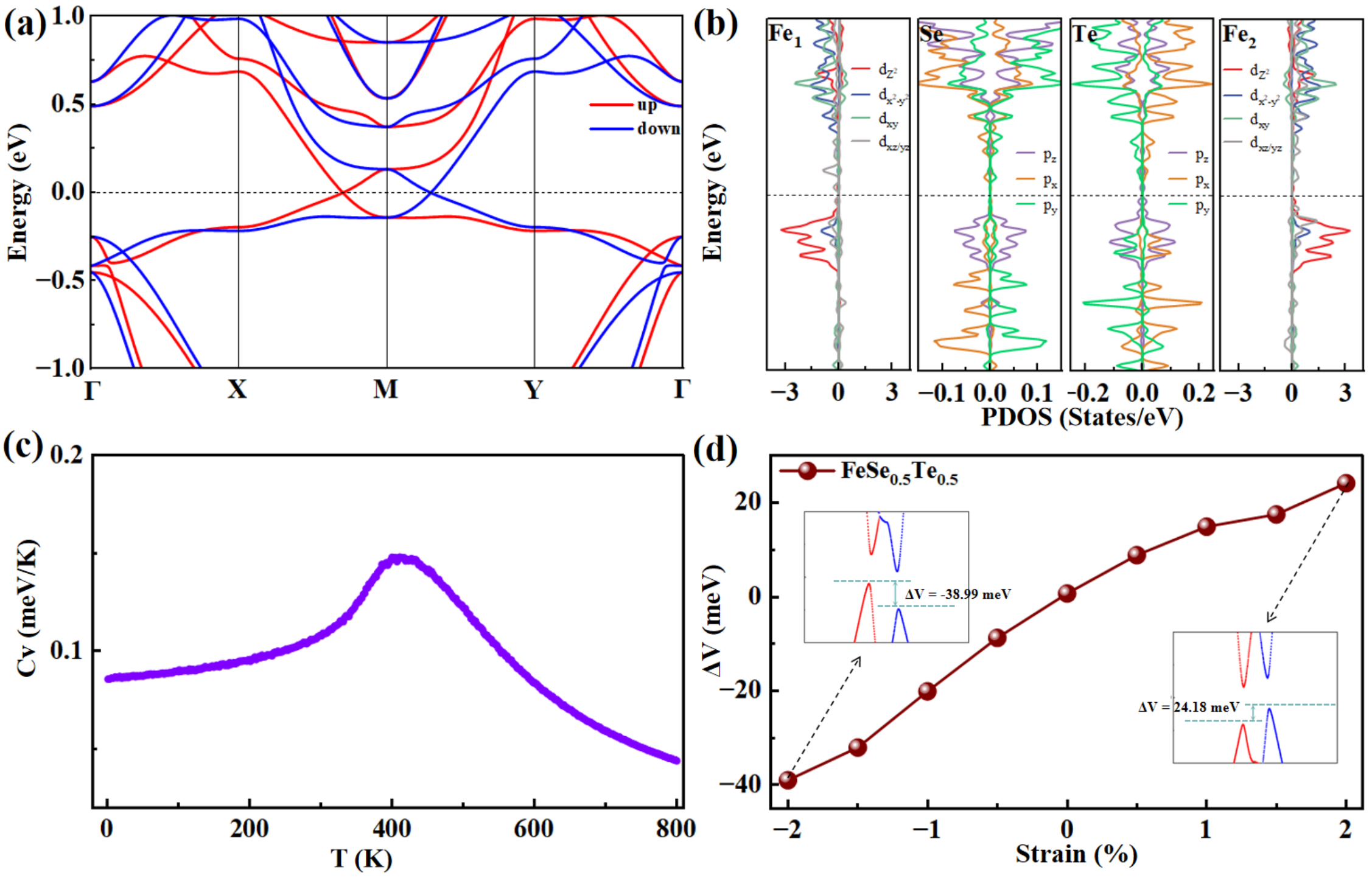

*Fig. 3. Properies of monolayer $FeSe_{0.5}Te_{0.5}$. (a) and (b) The band structure and PDOS. (c) Temperature dependence of specific heat. (d) The valley polarization gap $(\Delta V)$ as a function of in-plane strain.*

To further explore the stability of the altermagnetism, the corresponding Néel temperature was calculated by using Monte Carlo simulations based on the parameters extracted from the classical Heisenberg model:

$$H = H_0 - J_1 \sum_{<m,n>_{intra}} S_m \cdot S_n - J_2 \sum_{<m,n>_{intra}} S_m \cdot S_n \quad (4)$$

where, $J_1$ and $J_2$ represent the nearest and next-nearest neighbor exchange interaction parameters, respectively, $S_{m/n} = 3/2$ for Fe atoms based on the DFT calculations. Here, we computed the total energies of four distinct magnetic states of a 3×3 supercell (see Fig. S5). By fitting these energies to Eq. 4, the exchange interaction parameters were extracted and summarized in Table I. One can see that $FeSe_{0.5}Te_{0.5}$, $FeS_{0.5}Te_{0.5}$ and $FeS_{0.5}Se_{0.5}$ exhibit exceptionally large exchange interaction parameters. These strong exchange couplings give rise to high Néel temperatures ($T_N$) as shown in Fig. 3(c), Fig. S6(c) and Fig. S7(c). This is consistent with previous reports on FeSe-based magnetic systems [51]. Consequently, these robust superexchange interactions lead to elevated magnetic ordering temperatures for altermagnetic Janus $FeX_{0.5}Y_{0.5}$ monolayer systems.

Table I. The exchange interaction parameters and Néel temperatures for $FeSe_{0.5}Te_{0.5}$, $FeS_{0.5}Te_{0.5}$ and $FeS_{0.5}Se_{0.5}$, respectively.

| | $J_1$ (meV) | $J_{2a}$ (meV) | $J_{2b}$ (meV) | $T_N$ (K) |
|---|---|---|---|---|
| $FeSe_{0.5}Te_{0.5}$ | -43.17 | -52.09 | -62.61 | 415 |
| $FeS_{0.5}Te_{0.5}$ | -61.96 | -35.86 | -75.14 | 335 |
| $FeS_{0.5}Se_{0.5}$ | -86.25 | -46.55 | -79.66 | 360 |

Moreover, one can notice from the band structures in Fig. 3(a) that, there is significant spin splittings between the valleys near the M point along $\Gamma-\mathrm{X}$ and $\Gamma-\mathrm{Y}$ directions. This offers a valley-like degree of freedom and provides new opportunities for next-generation spintronic and valleytronic devices. To harness this valley-like degree, the key step is to break the population equilibrium between the valleys, i.e., achieving valley polarization [52]. Unlike conventional valleytronics, where the spin splitting stems from strong spin-orbital coupling, the valley-like degree of freedom in Janus $FeX_{0.5}Y_{0.5}$ is intrinsically interlocked by the crystal symmetry. Consequently, the corresponding valley polarization can be achieved by breaking the in-plane crystal symmetry through uniaxial strain [53,54]. From DFT calculations, one can see that the valley polarization gap $(\Delta V)$ can be tuned from -40 meV to 25 meV within a reasonable range of in-plane strain (-2 % to 2 %) as shown in Fig. 3(d). Notably, a nonzero net Hall conductivity can be obtained by shifting the Fermi level to intercept only one valley, for example via electrostatic gating or chemical doping. This selective occupation breaks the balance of anomalous velocity contributions from opposite valleys, thereby giving rise to a net nonzero Hall conductivity. Such a mechanism offers a simple and effective route to generate and control valley-polarized Hall conductivity without relying on external magnetic fields.

Although the altermagnetism here is fundamentally independent of the spin–orbit coupling (SOC), Janus $FeX_{0.5}Y_{0.5}$ possesses appreciable SOC, while monolayer FeSe and $Fe(Te_xSe_{1-x})$ have been confirmed as topological superconductors [55]. Consequently, it is imperative to investigate whether the altermagnetic $FeX_{0.5}Y_{0.5}$ possesses topological properties, as the interplay between them may give rise to fundamentally new quantum phenomena. DFT calculations show that a large SOC-induced band gap of 51.4 meV (38.3 meV, 47.1 meV) exists for $FeSe_{0.5}Te_{0.5}$ ($FeS_{0.5}Se_{0.5}$,

$FeS_{0.5}Te_{0.5}$) as shown in Fig. 4(a) (Fig. S8 and Fig. S9).

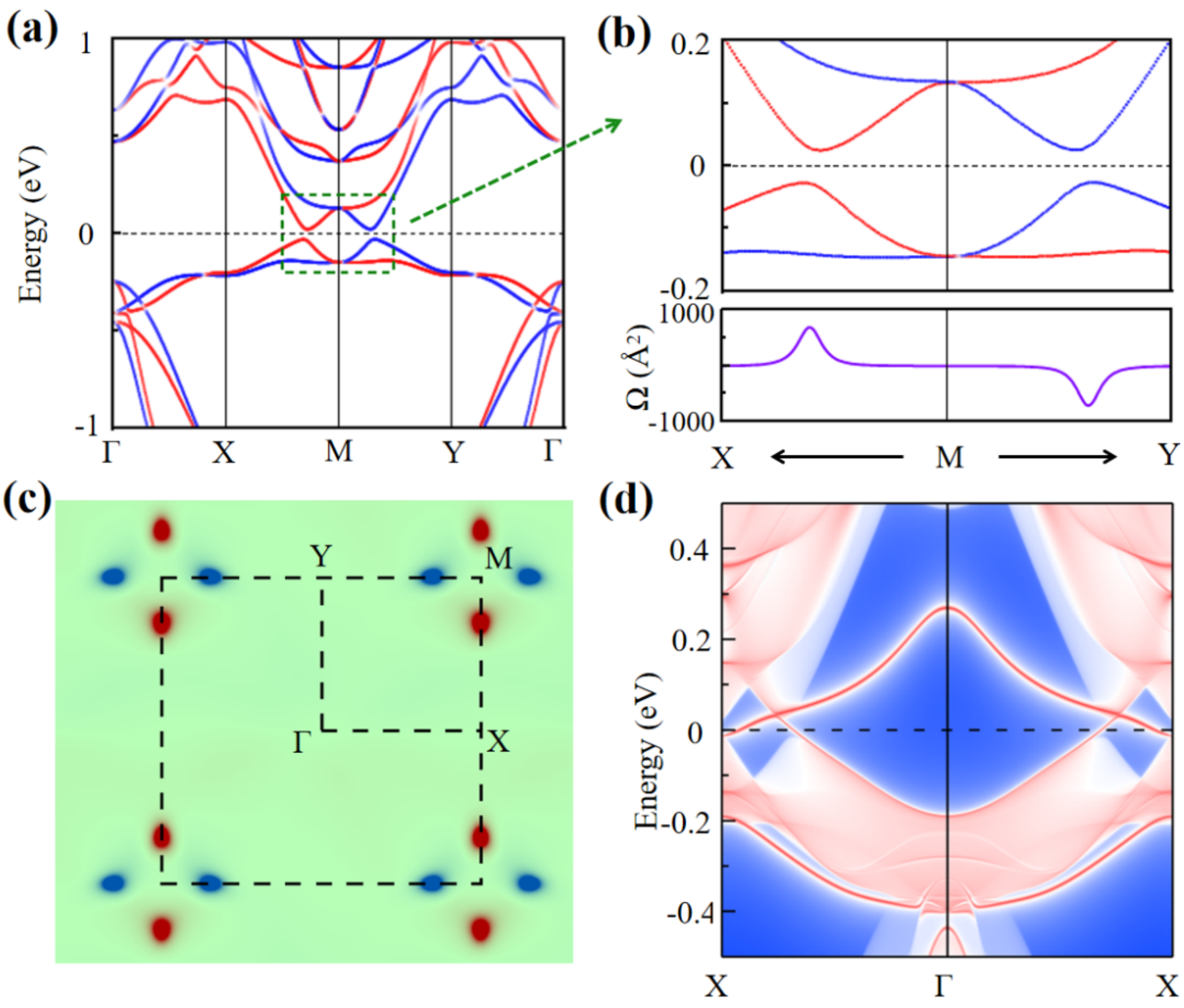

*Fig. 4. (a) The band structure of $FeSe_{0.5}Te_{0.5}$ with the spin-orbit coupling (SOC) included. (b) The zoomed-in band structure (upper panel) and the Berry curvature Ω (lower panel) along X–M–Y. (c) The Berry curvature distribution in the Brillouin zone of $FeSe_{0.5}Te_{0.5}$. (d) The corresponding one-dimensional band structure.*

These sizable gaps are a highly desirable feature for topological materials. Meanwhile, the Berry curvature Ω(k) along the high-symmetry X-M-Y path in $FeSe_{0.5}Te_{0.5}$ exhibits two pronounced peaks of opposite sign (see the bottom of Fig. 4(b)), which are directly associated with altermagnetic spin-valley polarization. Further analysis of the Berry curvature over the entire first Brillouin zone reveals four positive and four negative contributions localized near the zone corners as shown in Fig. 4(c). Consequently, the integration of Berry curvature over the entire Brillouin zone yields a vanishing net Hall conductance, a behavior of quantum spin Hall effect. To further confirm the intrinsic topological nature of $FeSe_{0.5}Te_{0.5}$, we calculate the spin Hall conductance ($\sigma_{xy}^{S}$) using a Wannier representation constructed from maximally localized Wannier functions (MLWFs) based on the first-principles calculations as shown in Fig. S10. We found $C_s = -1$ for $FeSe_{0.5}Te_{0.5}$, $FeS_{0.5}Se_{0.5}$, and $FeS_{0.5}Te_{0.5}$, indicating nontrivial spin topology in all three monolayers. We further construct a tight-binding (TB) model using parameters extracted from the DFT band structure and compute the electronic spectrum of $FeSe_{0.5}Te_{0.5}$ nanoribbons. The emergence of robust edge states across the Fermi level, as shown in Fig. 4(d), provides direct and compelling evidence of the nontrivial topological character of these materials. These results reinforce the potential of $FeX_{0.5}Y_{0.5}$ for applications in low-power spintronic and topological devices.

## IV. CONCLUSION

In summary, the DCB-lattice model is proposed to realize the spin-splitting electronic states in 2D altermagnetic iron-based superconductors. This altermagnetism is induced by introduction of anisotropic local environments for the two magnetic Fe sublattices via employing the Janus structure of $FeX_{0.5}Y_{0.5}$ with different types of chalcogen atoms at top and bottom atomic layers. Through first-principles calculations, the mechanism is demonstrated in Janus $FeX_{0.5}Y_{0.5}$ (X≠Y; X, Y = S, Se, Te) monolayers. DFT calculations show that these structures have high Néel temperature due to the strong exchange interactions and valley-like degree of freedom. We further reveal that valley polarization in monolayer $FeX_{0.5}Y_{0.5}$ can be effectively tuned via applied in-plane strain and the resulting valley-polarized anomalous Hall conductivity can be manipulated through shifting the Fermi level. In addition, these Janus $FeX_{0.5}Y_{0.5}$ monolayers have large topological band gaps. Our work suggests a new strategy for realizing and controlling spin-splitting electronic states in superconducting materials, and inspires further explorations on superconducting spintronics.

## ACKNOWLEDGMENTS

The authors thank prof. Hernández for informing us of their recently published related work [21] during the process of refining our manuscript. This work was partially supported by the National Natural Science Foundation of China (No. 12374046, 12304089, 12634002), the Shanghai Science and Technology Innovation Action Plan (Grant No. 24LZ1400800). Calculations were partially performed on computers at Shanghai Technical Service Center of Science and Engineering Computing, Shanghai University.

## DATA AVAILABILITY

The data that support the findings of this article are not publicly available upon publication because it is not technically feasible and/or the cost of preparing, depositing, and hosting the data would be prohibitive within the terms of this research project. The data are available from the authors upon reasonable request.

Rev. B **92**, 165104 (2015).